\documentclass[prd,twocolumn,nofootinbib]{revtex4}
\usepackage{latexsym}
\usepackage{amsthm}
\usepackage{amsmath}
\usepackage{graphicx}  
\usepackage{bm}        
\usepackage{amssymb}   
\usepackage{hyperref}
\newcommand{\gsim}{\lower.7ex\hbox{$\;\stackrel{\textstyle>}{\sim}\;$}}
\newcommand{\lsim}{\lower.7ex\hbox{$\;\stackrel{\textstyle<}{\sim}\;$}}

\newcommand{\be}{\begin{equation}}
\newcommand{\ee}{\end{equation}}
\newcommand{\bea}{\begin{eqnarray}}
\newcommand{\eea}{\end{eqnarray}}

\newcommand{\bef}{\begin{figure}[htbp]\begin{center}}
\newcommand{\eef}{\end{center}\end{figure}}

\begin{document}
\title{Twisting loops and global momentum non-conservation in Relative Locality}
\author{Andrzej Banburski \\
\textit{Perimeter Institute for Theoretical Physics, \\
31 Caroline St. N, Waterloo, ON N2L 2Y5, Canada}}

\begin{abstract}
Recent work in Relative Locality has shown that the theory allows for a solution of an on-shell causal loop. We show that the theory contains a different type of a loop in which locally momenta are conserved, but there is no global momentum conservation. Thus a freely propagating particle can decay into two particles, which later recombine to give a particle with momentum and mass different than the original one.
\end{abstract}

\maketitle

\section{Introduction}

Relative Locality (RL) was proposed in \cite{AmelinoCamelia:2011bm} as a framework to study the dynamics of interacting point particles in presence of non-trivial geometry of momentum space. There is evidence from $2+1$-dimensional Quantum Gravity \cite{Freidel:2003sp, Schroers:2007ey} that when one considers point particles, the resulting goemetry of momentum space is deformed by Planck-scale effects. RL is thus thought to be the limit of Quantum Gravity in which $\hbar , G \rightarrow 0$, but $m_{Planck} = \sqrt{\frac{\hbar c}{G}}$ stays fixed. In this peculiar (not fully classical) limit, one can expect there to be some remnants of the quantum behaviour of particles. Specifically, it is not completely obvious whether the theory allows for loops that are not classically allowed in Special Relativity, but are allowed at the quantum level. The inclusion of the invariant Planck scale might lead to even more interesting phenomena. For example, it was shown in \cite{Freidel:2011mt} that two photons of different energies emitted simultaneously arrive to the same place at different times. This has been proposed as a possible explanation of the observed time delay of photons arriving from Gamma Ray Bursts.

Recently it was shown in \cite{Chen:2012fu} that the theory posseses an on-shell causal loop solution. Motivated by this one might want to check whether there are more on-shell loop processes that are not allowed in Special Relativity, but are allowed in RL. We find that indeed the causal loop is not the only such solution. Surprisingly, it is possible to construct a twisting process in which a particle $p$ decays into two particles (call them $k$ and $l$), which after propagating for some time recombine to give a particle $q$ with momentum different than the original particle $p$, meaning that total momentum is not conserved. This process is obviously not allowed in Special Relativity, as you cannot embed it in Minkowski spacetime. In RL however, momentum space is taken as a primary object, so there is no universal notion of spacetime. If one exists, it is emergent as a mapping between the momentum space and the history of a net of processes. 

We first study this process in the most studied example of deformations of Poincar\' e symmetry, the $\kappa$- Poincar\' e \cite{Gubitosi:2011ej}, which corresponds to a momentum space with a torsionful and non-metric flat connection. In this case the process leads to change of mass of the particle. We find that if one allows for such twisting loops in the theory, then evaluating a process at 0 initial momentum and transforming to rest frame are not equivalent statements. 

We then investigate what happens in the case of recently studied Snyder momentum space \cite{Banburski}, in which metricity is present. It turns out that the process is still a solution, but the masses of the particles do not change. We finish with the discussion of the results and what they mean for Relative Locality.

\section{Relative Locality in a Nutshell}
In this section we will review the basic structure of Relative Locality. For more details on any of the topics, see the original paper \cite{AmelinoCamelia:2011bm}. We will describe the dynamics of particles in momentum space, without assuming what kind of geometry this space posseses - it should be ascertained experimentally.

In RL the notion of mass acquires a geometric meaning - it is the geodesic distance from the origin to point $p$, i.e. a particle of momentum $p$ has the mass given by
\begin{equation}
D^2(p) = m^2.
\end{equation}
Measuring the momenta and masses of particles allows us to reconstruct the metric $g^{\mu\nu}$ on the momentum space.

The other crucial notion is that of addition of momenta, which in general does not have to be that of vector addition. In generic case, the rule of composition of momenta does not necessarily have to be commutative nor associative. Let us define this operation as
\begin{equation}
\begin{split}
\oplus &: \mathcal{M}\times\mathcal{M} \rightarrow \mathcal{M} \\
&(p,q)\mapsto p\oplus q
\end{split}
\end{equation}
such that it has an identity $0$ 
\begin{equation}
0\oplus p = p\oplus 0 = p
\end{equation}
and an inverse
\begin{equation}
\ominus p \oplus p = p\ominus p = 0.
\end{equation}
It is useful to define a ``translated addition", by choosing some point $k$ as the new origin by which one defines addition:
\begin{equation}
p\oplus_kq = k\oplus\left(\left(\ominus k \oplus p\right)\oplus\left(\ominus
k\oplus q\right)\right).
\end{equation}

From this addition rule we can find the different properties of the geometry of our momentum space.
The connection is given by
\begin{equation*}
\Gamma^{\mu\nu}_\rho(p) = - \partial_r^\mu \partial_q^\nu\left(r\oplus_p q\right)_\rho \bigg|_{r = q = p}. 
\end{equation*}
If the connection is metric compatible, then the symmetric part of this gives us the Christoffel symbols, while the antisymmetric part measures the non-commutativity of the 
addition, thus gives the torsion
\begin{equation*}
T^{\mu\nu}_\rho(p) = -\partial_r^\mu \partial_q^\nu\left(r\oplus_p q - q\oplus_p r\right)_\rho \bigg|_{r = q = p}.
\end{equation*}
If the connection is not metric compatible however, we can define a nonmetricity tensor as
\begin{equation}
N^{\mu\nu\rho} = \nabla^\mu g^{\nu\rho}.
\end{equation}

One can also find that the nonassociativity of $\oplus$ leads to the curvature
\begin{equation*}
R^{\nu\alpha\beta}_\mu\!(p) \!=\! 2\partial_r^{\left[\nu\right.}\partial_q^{\left.\alpha \right]}\partial_k^\beta
\!\left(\!\left(r\!\oplus_p\! q\right)\!\oplus_p\! k\! -\! r\!\oplus_p\!\left(q\!\oplus_p \!k\right)\!\right)_\mu \! \bigg|_{r=q=k=p}
\end{equation*}

We can now write an action for the theory with $N$ particles and $M$ interactions
\begin{equation}
S = \sum_{J=1}^{N}S^J_{\textnormal{free}} + \sum_{ i=1}^{M} S^i_{\textnormal{int}}. \label{eq:action}
\end{equation}
The free part of the action is
\begin{equation}
S_{\textnormal{free}}^J = \int \textnormal{d}s \left(x^\mu_J \dot{p}_\mu^J + \mathcal{N}_J\left(D^2\left(p\right) - m^2\right) \right),
\end{equation}
where $ \mathcal{N}_J$ is a Lagrange multiplier for the mass-shell constraint. Note that $x_J^\mu$ is just a Lagrange multiplier enfrocing the conservation of momentum for freely propagating particle, and is formally an element of the cotangent space of the momentum manifold, $x_J^\mu \in T^*_{p^J}\mathcal{M}$. The notion of spacetime can emerge if we create a net of relations between the different $x$'s for different particles. For this we cave to consider the interactions, which are given by the action
\begin{equation}
S_{\textnormal{int}}^i = z_i^\mu\mathcal{K}^i(p(s_i))_\mu,
\end{equation}
where $\mathcal{K}$ is a conservation law for momenta (for example $\mathcal{K} =  p\oplus q \ominus r = 0$) and $s_i$  is the time of this interaction. $z$'s are again just Lagrange multipliers, but can be considered to be some sort of ``interaction coordinates".

By varying the action, we can easily get
\begin{equation}
\begin{split}
\delta S = \sum_{J=1}^{N}\int_{s_{i_1}}^{s_{i_2}}\bigg(\delta x^\mu_J \dot{p}_\mu^J - \left[\dot{x}^\mu_J - \mathcal{N}_J\frac{\delta D^2(p)}{\delta p_\mu^J}\right] \\ + \delta\mathcal{N}_J \left(D^2(p^J)-m^2_J\right) \bigg) + \textnormal{boundary terms}
\end{split}
\end{equation}

From this variation we get the equations of motion along a worldine of a particle (that is excluding the endpoints) to be 
\begin{equation}
\begin{split}
\dot{p}^J_\mu &= 0 \\
D^2(p^J) &=  m_J^2 \\
\dot{x}_J^\mu &= \mathcal{N}_J\frac{\delta\left(D^2(p^J) - m^2_J\right)}{\delta p^J_\mu}
\end{split}
\end{equation}
There is nothing really surprising about these equations of motion. The only non-standard looking one is third equation, which is just the definition of 4-velocity.
The more nontrivial relation comes from the variation of the boundary terms:
\begin{equation}
\delta S_\textnormal{bdry}=\sum_{i=1}^M \mathcal{K}^i_\mu \delta z_i^\mu - \left(x_J^\mu (s_i) \pm z^\nu_i \frac{\delta \mathcal{K}^i_\nu}{\delta k_\mu^J}\right)\delta k_\mu^J(s_i)
\end{equation}
where sign depends on whether the particle is incoming or outgoing in the interaction. We have then
\begin{equation}
\begin{split}
\mathcal{K}_i=0 \\
x^\mu_J(s_i) &=\pm z^\nu\frac{\delta\mathcal{K}_\nu}{\delta p^J_\mu}.\label{eq:rleom}
\end{split}
\end{equation}
The first equation is obvious. The other one however, is very non-trivial - this is the equation that tells us how spacetime emerges from connections between the different $x$ points in a net of events.

\section{Twisting loop}
\begin{figure}[h]
	\centering
		\includegraphics[width=0.30\textwidth]{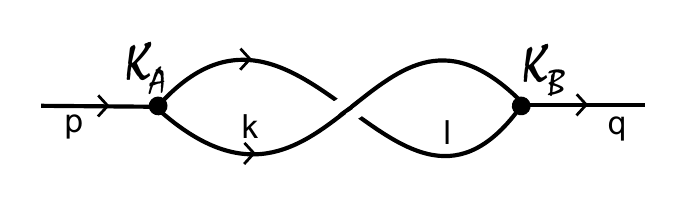}
	\caption{A twisting loop, with $p \neq q$}
	\label{fig:loop}
\end{figure}

Let us consider the simple set of conservation laws\footnote{A different set of conservation laws might lead to different physics if the addition rule is non-commutative or non-associative.}
\begin{equation}
\mathcal{K}_A = p\ominus \left(k\oplus l\right), \ \ \ \ \   \mathcal{K}_B = \left(l\oplus k\right)\ominus q \label{eq:conservation}
\end{equation}
This set of conservation laws describes a twisting loop process shown in Figure \ref{fig:loop}. 

To impose that the two particles $k$ and $l$ meet at the interaction events $\mathcal{K}_A$ and $\mathcal{K}_B$, we have to make them satisfy all of the equations of motion. To write the condition we want to satisfy, let us define two transport operators which are mappings $T^*_k\mathcal{M}\rightarrow T^*_l\mathcal{M}$:
\begin{equation}
\begin{split}
\left[M_A\right]_\mu^\nu \equiv & \left(-\frac{\partial \mathcal{K}_{A \rho}}{\partial l_\nu}\right) \cdot \left(-\frac{\partial \mathcal{K}_{A \rho} }{\partial k_\mu}\right)^{-1} \\
\left[M_B\right]_\mu^\nu \equiv & \left(\frac{\partial \mathcal{K}_{B \rho}}{\partial l_\nu}\right) \cdot \left(\frac{\partial \mathcal{K}_{B \rho} }{\partial k_\mu}\right)^{-1}
\end{split}
\end{equation}
Notice that by Eq. (\ref{eq:rleom}), these operators are the unique ones corresponding to the physical process described by conservation laws (\ref{eq:conservation}).

We now have to impose that the particles $k$ and $l$ propagate forward in time in their respective phase spaces in such a way that the third and fourth equations of motion are satisfied. We require that the endpoints of the two worldlines are related to each other by the two interaction coordinates $z_A$ and $z_B$. Let $x$ denote the coordinates on the worldine of one of the particles inside the loop. We then require that $x_B = x_A + \dot{x} \tau$, where the labels $A,B$ refer to the two endpoints. We also have that at the interaction the two endpoints of the two interactions are related to each other by the last of the equations of motion (\ref{eq:rleom}). We can express these conditions in the single equation
\begin{equation}
\left[M_B\right]_\mu^\nu \left(x_{k,A}^\mu + u_k^\mu \tau_k\right) = \left[M_A\right]_\mu^\nu x_{k,A}^\mu + u_l^\nu \tau_l \label{eq:condition}
\end{equation}
where $x_{k,A} \in  T^*_k\mathcal{M}$ is the location of the event A on the Hamiltonian spacetime of the particle $l$, $u_{k, l}$ are the velocities of the particles $k$ and $l$ respectively, and $\tau_ {k,l}$ are the proper times for the propagation of $k$ and $l$. Again, this is a unique condition that has to be satisfied in order for the processes to form a loop.

Note that this looks very similar to the condition for the existence of a causal loop in \cite{Chen:2012fu}, with the important difference that there is a relative minus sign between the two terms with four-velocities. This will allow us to have much more solutions.

If the conditions (\ref{eq:conservation}) and (\ref{eq:condition}) are satisfied, then we have found a solution to the theory. Indeed, as we will now show, these can be easily satisfied. It is important to point out, that unlike in the causal loop case, this loop can happen at $x=0$, because of the mentioned difference in sign between the two velocities. Nonetheless, the solutions of this set of equations are going to to have the property of ``x-dependence" which was discussed in \cite{Chen:2012fu}. This means that translation invariance is explicitly broken here, as the set of proper times and momenta solving the conditions  (\ref{eq:conservation}) and (\ref{eq:condition}) depends on the specific cotangent space coordinate $x$. We will discuss this point more in the last section.

We will now solve these equation in the specific case of the geometry of $\kappa$-Poincar\'e, which has been one of the most studied examples in DSR - Doubly (or Deformed) Special Relativity \cite{AmelinoCamelia:2000ge, Magueijo:2002am}, a precursor of RL. As a geometry for Relative Locality it has been studied in \cite{Gubitosi:2011ej}. $\kappa$-Poincar\'e is one of the weakest deformations of Poincar\'e group to a Hopf algebra \cite{Majid:1994cy}, in which only boost sector is deformed, and the deformation scale is $\kappa$, which for phenomological reasons is usually taken to be Planck scale. 

It was shown in \cite{Gubitosi:2011ej} that $\kappa$-Poincar\'e is described by de Sitter space of radius $\kappa$ with the metric 
\begin{equation}
ds^2 = dE^2 - e^{2E/\kappa}dp^2.
\end{equation}

The deformation of the boosts leads to the deformed addition rule for momenta, given by
\begin{equation}
\begin{split}
\left(p\oplus q\right)_0 &= p_0 + q_0 \\
\left(p\oplus q\right)_i &= p_i + e^{-p_0/\kappa}q_i .
\end{split}
\end{equation}
From this one finds that the connection is not metric compatible, and has non-vanishing torsion. It is easy to see however that this addition rule is associative, and thus leads to no curvature. 

The mass-shell constraint (found from the geodesic distance) is given by 
\begin{equation}
m = \kappa \  \textnormal{cosh}^{-1}\left(\cosh\left(p_0/\kappa\right) - e^{p0/\kappa}\frac{|\vec{p}|^2}{2 \kappa^2}\right).
\end{equation}

We will proceed to solve the equations (\ref{eq:conservation}) and (\ref{eq:condition}) using the above properties of $\kappa$-Poincar\'e momentum space.
The general solution is quite complicated, so we focus on the case of $x=0$. We can solve the constraints and for example find the expressions for $\tau_l$ and $k_i$ in terms of $l_0, l_i$ and $k_0$. We get that $\tau_k = \tau_l$ and
\begin{equation}
k_i = e^{-k_0} l_i \frac{\sinh m_k}{\sinh m_l}.
\end{equation}
Note however, that this expression is really an equation for $k_i$ as mass depends on the evalue of the momentum. More explicitly we have
\begin{eqnarray}
k_i = -\frac{e^{- k_0 - l_0} l_i \left(e^{k_0}+e^{ k_0+2 l_0} \left(-1+l_i^2\right)+F \right)}{2 l_i^2} \ \ \ \ \  \\
F \!=\! \sqrt{\!e^{2 k_0}\!+\!4 e^{2 l_0} l_i^2\!+\!e^{2 k_0+4 l_0}\! \left(l_i^2 \!-\!1\right)^2 \!-\! 2 e^{2 k_0+2 l_0}\! \left( l_i^2\!+\!1\right)}\nonumber
\end{eqnarray}

\begin{figure}[h]
	\centering
		\includegraphics[width=0.30\textwidth]{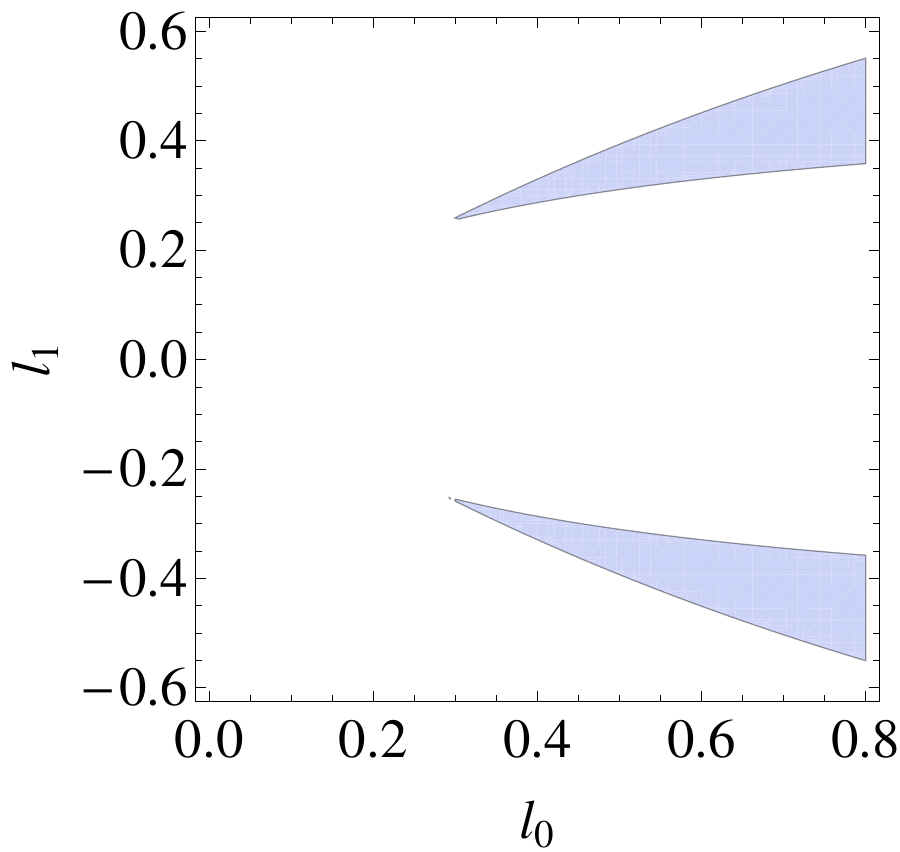}
	\caption{Region of the plot for which there are solutions, with $k_0 = 0.2$.We restricted the region by requiring that mass of $p$ be at least $3\%$ higher than that of $q$. The highest mass with this choice is at least $10\%$ higher. It is not obvious how much mass difference is possible in general, though numerics show it to be at least $12\%$. Without the restriction on big mass increase, the region of solutions is nearly the whole light-cone.}
	\label{fig:region}
\end{figure}

This allows us finally to investigate in which region we have physical solutions with $k$ and $l$ non-zero. It is crucial that we impose the momentum conservation laws and that the resulting momenta $p$ and $q$ satisfy the mass-shell condition. We plot the region in Figure \ref{fig:region}.

We find that in general $p \neq q$. Note however, the linear addition of energies in $\kappa$-Poincare means that only the spatial components are different. This means that the masses of the two particles can be different.

An example solution is given by the set of momenta:
\begin{equation}
\begin{split}
l_0 = 0.2, \ \ l_1 = 0.1, \ \ k_0 = 0.2, \ \ k_1 \approx 0.086, \\  
p_0 = q_0 = 0.4, \ \ p_1 \approx 0.168, \ \ q_1 \approx 0.171 
\end{split}
\end{equation}
In this case we have $m_p \approx 1.005 \  m_q$. 

We could try to understand this result better by going to rest frame. However, it was pointed out in \cite{Gubitosi:2011ej} that boosts are modified in $\kappa$-Poincare momentum space. In 1+1D, a $\kappa$-boosted momentum $p$ is given by
\begin{equation*}
\left( \begin{array}{c}\!\!\! p_0 \!+\! \kappa\! \log\! \left[ \left( \mathrm{ch} \xi/2  \!+\! \frac{p_1}{\kappa} \mathrm{sh} \xi/2 \right)^2 \!-\! e^{- 2 p_0 / \kappa} \mathrm{sh}^2 \xi/2  \right]\!\!
\\
\kappa \frac{\left( \mathrm{ch} \, \xi/2 + \frac{p_1}{\kappa} \mathrm{sh} \, \xi/2 \right) \left( \mathrm{sh} \, \xi/2 + \frac{p_1}{\kappa} \mathrm{ch} \, \xi/2 \right) - e^{-2 p_0/\kappa} \mathrm{ch} \, \xi/2 ~ \mathrm{sh} \, \xi/2 }{\left( \mathrm{ch} \, \xi/2 + \frac{p_1}{\kappa} \mathrm{sh} \, \xi/2 \right)^2 - e^{- 2 p_0 / \kappa} \mathrm{sh}^2 \, \xi/2 }  \end{array} \right) 
\end{equation*}
where $\xi$ is rapidity, sh stands for $\sinh$ and ch for $\cosh$. Call this transformed momentum $\Lambda\left(\xi, p\right)$. It turns out however that for addition of momenta we have
\begin{equation}
\Lambda( \xi,p\oplus q) \neq \Lambda(\xi,p) \oplus \Lambda(\xi,q) ~.
\end{equation}
Instead we have backreaction of one of the momenta on the rapidity. Defining
\begin{equation*}
\xi \triangleleft p \!=\! 2 \, \mathrm{arcsinh} \!\left( \frac{e^{-p_0/\kappa} \sinh \frac{\xi }{2} }{\sqrt{\!\left( \cosh \!\frac{\xi }{2}  \!+\! \frac{p_1}{\kappa} \sinh \!\frac{\xi }{2} \right)^2 \!\!\!-\! e^{-2 p_0/\kappa} \sinh^2 \frac{\xi }{2}}} \right) 
\end{equation*}
we have that for a composition rule, the $\kappa$-boost is given by\footnote{Our analysis here is valid in our specific 1+1D example. In 3+1D a coproduct of boosts in $\kappa$-Poincar\'e contains a rotation, see  \cite{Gubitosi:2011ej}. This however, does not influence the results that follow.} 
\begin{equation}
\Lambda (\xi, q \oplus k ) =
\Lambda (\xi, q) \oplus \Lambda (\xi \triangleleft q, k ).
\end{equation}
 With this, we can find a $\xi$ such that $p_1 = 0$. Our process in the $\kappa$-boosted frame is
\begin{equation}
\Lambda\left(\xi, p\right) \rightarrow \Lambda\left(\xi, k\oplus l\right) \rightarrow \Lambda\left(\xi, l\oplus k\right) \rightarrow \Lambda\left(\xi, q\right)
\end{equation}
Note, however, that under a Lorentz transformation the twisting of the particles $k$ and $l$ means that for one vertex we have backreaction on rapidity from $k$ and in the other from $l$. This is one reason for which intuition breaks down in relativity of locality of $\kappa$-Poincar\'e.

Numerically we have for $\xi \approx -0.664109$
 \begin{equation}
\begin{split}
l_0 ' &\approx 0.168, \ \ l_1 ' \approx 0.017, \ \ k_0 ' \approx 0.177, \ \ k_1 ' \approx -0.015, \\  
q_0 ' &\approx 0.343, \ \  q_1 ' \approx 0.003, \ \  p_0 ' \approx 0.345, \ \ p_1 ' = 0
\end{split}
\end{equation}

The picture we get then is that a static particle decays into $k$ and $l$, which head in opposite directions, and which later recombine to give $q$ with different mass than $p$'s and with kinetic energy. This process is a solution to the theory, and it can happen because in Relative Locality we have abandoned the notion of a universal spacetime.

Applying the $\kappa$-boost to the conservation laws, we get
\begin{equation}
\begin{split}
\mathcal{K}_A' &= \Lambda (\xi, p) \oplus \Lambda (\xi \triangleleft p, \ominus k ) \oplus \Lambda (\xi \triangleleft p\ominus k, \ominus l ) \\
\mathcal{K}_B' &= \Lambda (\xi, l) \oplus \Lambda (\xi \triangleleft l,  k ) \oplus \Lambda (\xi \triangleleft l\oplus k, \ominus q )
\end{split}
\end{equation}
For the Eq. (\ref{eq:condition}) we use the Lorentz transformation at the $\mathcal{K}_B$ vertex to find the new 4-velocities. Under the deformed Lorentz transformation, the equations of motion change in such a way, that the transformed solution is still a solution to the theory. In this manner, the solution is Lorentz invariant. It is strange however, that the solution for $p_i = 0$ in the lab frame is not physically equivalent to having a nontrivial solution in this frame and boosting it to the rest frame of $p$. Perhaps, if we allow for such twisting graphs (and the equations of motion do allow it), then the notion of ``going to rest frame" is not necessarily equivalent with ``evaluating the process at 0 momentum". The reason for this is perhaps the non-equal treatment of momenta under $\kappa$-boosts of composition rule in the geometry of $\kappa$-Poincar\'e. This assymetry of treatment of the two particles is also present when applying the boosts to their free propagation. At the algebraic level, this strange behavior of boosts comes from the fact that the $\kappa$-boosts do not form a subalgebra within the $\kappa$-Poincar\'e Hopf algebra. Thus, when the translation invariance is broken (because our solution depends on the specific choice of $x$) it should not be so strange that our intuition of boosting to a rest frame is lost.

\section{Mass conservation in Snyder momentum space}
Let us now consider the same process in the recently studied geometry of Snyder momentum space \cite{Banburski}. Snyder spacetime \cite{Snyder:1946qz} was the first example of quantized spacetime \cite{Doplicher:1994tu} ever considered. Snyder showed that if one allows the spacetime coordinates to become Hermitian operators, then the resulting theory can be made Lorentz invariant and at the same time posses a minimal length scale. The construction of this spacetime requires a non-trivial geometry of momentum space, namely the Snyder momentum space is a de Sitter space with a metric compatible connection. Thus this momentum space is an exact opposite of $\kappa$-Poincar\'e, in that it does not have nonmetricity or torsion. de Sitter space (dS) can be described as a hyperboloid in 5d Minkowski space. If $p$ is a point on dS, then the Mikowski coordinates $P_A(p)$, $A=0, \ldots , 4$, are constrained by
\begin{equation}
\eta^{AB} P_A P_B = \kappa^2,
\end{equation}
where $\kappa$ is the radius of curvature. The addition rule for momenta is constructed so that
\begin{equation}
\Lambda\left(Q\oplus K\right) = \Lambda\left(Q\right)\oplus\Lambda\left(K\right),
\end{equation}
where $\Lambda \in \textnormal{SO(3,1)}$ is a Lorentz transformation. We have
\begin{equation}
\begin{split}
(P\oplus Q)_4 &= \frac{2P_4 Q_4 - P\cdot Q}{\kappa} \\
(P\oplus Q)_\mu &= Q_\mu + P_\mu\frac{\kappa Q_4 + 2P_4 Q_4 - P\cdot Q}{\kappa^2+P\cdot I}.
\end{split}
\end{equation}
and importantly the 4th component is related to the mass of the particle by $P_4^2 = \kappa^2 \cosh \left(m/\kappa\right)$. Thus the addition of the 4th component is just a deformed addition of masses. By the virtue of the construction, this addition rule leads to nonvanishing curvature, but vanishing torsion and nonmetricity tensors.

It is easy to notice that in Snyder momentum space $(P\oplus Q)_\mu \neq (Q\oplus P)_\mu$. However, for the 4th component, $(P\oplus Q)_4 = (Q\oplus P)_4$, and so we immediately see that if our construction is a solution in this case, then the mass of $p$ and $q$ are necessarily the same.

The dependence on $x$ of  Eq. (\ref{eq:condition}) does not vanish in this case, so we can choose some specific point $x$ and get an ``x-dependent" solution to equations of motion. It is difficult to solve the equation in general, but the expressions simplify greatly in Snyder, when one considers massless particles. For this reason, let us make an ansatz (we will work with $\kappa = 1$):
\begin{equation}
k_4 = l_4 = 1, \ \ \ k = \left(k_0,k_0,0,0\right), \ \ \ l=\left(l_0,-l_0,0,0\right)
\end{equation}
We thus restrict ourselves to two dimensions. Setting $x_3=x_4=0$ and normalizing the 4-velocities to be $(1,\pm 1,0,0)$, we get that a solution to Eq. (\ref{eq:condition}) is
\begin{equation}
\begin{split}
\tau_l &= -\frac{ k_0 \left(k_0 (2 \tau_k +x_0+x_1)+\sqrt{\Delta}\right)}{4} \\
l_0 &= \frac{k_0 (2 \tau_k + x_0 + x_1)-\sqrt{\Delta}}{2 (x_0-x_1)}\\
\Delta &= 8 \tau_k (x_0-x_1)+k_0^2 (2 \tau_k+x_0+x_1)^2.
\end{split}
\end{equation}
This obviously has many solutions (though not $x=0$), one of which is for example for $x/\tau_k=(-39,-13,0,0)$:
\begin{equation}
\begin{split}
k_0 &= 0.3, \ l_0 \approx 0.37, \ \tau_k = 1, \ \tau_l \approx 0.82, \\
p_0 &\approx 0.7, \ p_1\approx -0.03, \ q_0 \approx 0.71, \ q_1\approx -0.11
\end{split}
\end{equation}
with mass $m_p=m_q \approx 0.653$ in units of $\kappa=1$. It is interesting that in this geometry the two propagation times are no longer equal (this might be because we chose $x \neq 0$). Note that this is a very high energy solution, but one can easily find a solution for any energy. In that case however, similarly with the causal loop, $x$ has to be far from the origin.

We get that again in general $p\neq q$, but importantly the mass does not change. We can thus form a loop in a theory where only two types of particles (one of them even massless) are present. Here imposing a restriction on masses allowed in the theory is not so easy to satisfy. It would seem that allowing such twisting loops in the theory with massive particles means that we have local conservation of momentum, but globally the momentum is not conserved.

It is important to note that here no issues arise with Lorentz transformations, as unlike in $\kappa$-Poincar\'e, the boosts are not deformed and form the group SO(3,1). This means that the issue of going to rest frame not being equivalent with evaluating a process at 0 initial spatial momentum is purely the result of deforming the boosts in $\kappa$-Poincar\'e geometry.

An interesting simplification occurs in Snyder momentum space if we restrict our attention to the theory with only massless particles. In that case we have to require that $k\cdot l = 0$. This can be satisfied by several 4-vectors, but for the sake of simplicity we will concern ourselves with a two dimensional case of $k=(k_0,k_0)$ and $l=(l_0,l_0)$. In general it can be shown that the addition simplifies to usual addition in Minkowski space and only nontriviality stays in the matrices $M_A$ and $M_B$. It is easy to show that the general solution of this problem is
\begin{equation}
\tau_k = \tau_l, \ \ \ l_0 = -\frac{k_0 (x_0+x_1)}{2 \tau_k+x_0+x_1}.
\end{equation}
One of the solution of this is for example
\begin{equation}
\begin{split}
\tau_k &= \tau_l = 1, \ \ x_0=x_1=-\frac{1}{2}, \ \ k_0=l_0 = 0.2, \\
p_0&=p_1=q_0=q_1 = 0.4.
\end{split}
\end{equation}
Hence with massless particles we do recover total momentum conservation, though it seems that the loop is still a solution according to equations of motion. One might argue that it is not possible to embed this in spacetime since we are working with particles travelling in one direction only, but this construction holds in more than 2 dimensions. For all practical purposes though this solution is identical with free propagation of the original particle.

\section{Discussion}
We have shown that in Relative Locality the decay-recombination process $p\Rightarrow k\oplus l \Rightarrow l\oplus k\Rightarrow q$ is allowed and that in general 
$p \neq q$. This means that allowing twisting loops in Relative Locality leads to non-conservation of total momentum. This is perhaps not so strange, as even in General Relativity there is no total momentum conservation if the metric of spacetime does not possess enough symmetries.  However, it is crucial to notice that here the masses of $p$ and $q$ are different if the geometry of the momentum space has non-vanishing nonmetricity tensor. We have shown that if the connection is metric compatible, then the masses of $p$ and $q$ are the same. This means that a freely propagating massive particle in the theory might suddenly ``decide" to change its mass (in the case of nonmetricity) and momentum by such a loop process.

Similarly as in \cite{Chen:2012fu}, the issue of ``x-dependence" arises here. This means that the specific process with given proper times and momenta can only happen at some specific point on the Hamiltonian spacetime. This explicitly breaks translation invariance. In our case however, at least in the $\kappa$-Poincar\'e geometry, there is a solution at the origin $x=0$ of a cotangent plane $T^*_k\mathcal{M}$. It is not completely obvious if there exist solutions for all possible values $x$.

We have also found that in the case of $\kappa$-Poincar\'e momentum space, the solution is invariant under the deformed boosts, in the sense that a solution of the theory is transformed into a solution with the same physics. However, because of the backreaction of momenta on the rapidity for boosts acting on a sum of momenta, we have found that for our twisting loop, the statements of  ``going to rest frame" and ``evaluating the process at 0 momentum" are not neccessarily equivalent. This is because the boosts in $\kappa$-Poincar\'e do not form a closed subalgebra, which becomes apparent when translation symmetry is broken (``x-dependence").

The above discussion points towards several different conclusions, the simplest being that perhaps  graphs with twisting should not be allowed in the theory. First of all, it is not obvious what kind of spacetime structure would be necessary for description of such a process. Furthermore, what is responsible for the twisting? Should this be a dynamical process, or does it correspond to some micro-casual structure of the theory? The theory as described by the action Eq. (\ref{eq:action}) does not tell us which of the two it is. Perhaps an additional constraint, one that enforces momentum conservation globally, should be added to the action.

Another resolution of this problem is that perhaps only certain masses should be allowed in the theory, as we know that the spectrum of masses in our world is discrete. This route would perhaps exclude some undesirable loops solutions, but it is not obvious that one could not fit Standard Model masses to these loops. Also, as a mass superselection rule candidate, to find the "physically sensible" masses for particles, one would have to consider an infinite number of loop solutions. This is far from for example the clean predictions of the Randall-Sundrum model \cite{Randall:1999vf}. Notice that in the Snyder case we have shown that it is possible to construct the solution with only two types of particles, one being massless, thus making this solution not an easy way out.

 Also, if one were to consider quantizing Relative Locality, then there seems to be no obvious way to avoid these total momentum changing loops. A naive approach of considering particles with Bose statistics would require summing over all conservation laws for enforcing exchange symmetry, which would necessarily lead to the kind of twisting process described in this paper. To remove the kind of loops considered here, one requires a global description of processes, not a local one, so it would be difficult to construct a local QFT of Relative Locality preserving global momentum conservation. This might be not a problem, as one should perhaps expect some properly generalized notion of locality for the QFT of Relative Locality. As we noted before, in General Relativity for metrics without symmetries there is no global momentum conservation, and as RL is considered to be a limit of Quantum Gravity, we should not necessarily expect naive translation invariance to be a valid symmetry in the regime described by RL. However, we should still have a consistency of describing the same phenomenon by two different observers.

A possibly more promising avenue is that this loop constrains the geometries that should be allowed in the theory. As we have seen, in the case of Snyder momentum space only the total momentum was not conserved. In $\kappa$-Poincar\'e case though, the masses of the particles were changing and the Lorentz transformations lost their property of associating rest frame with evaluating the process at 0 momentum. This is perhaps a reason to constrain Relative Locality to geometries with a metric compatible connection, as nearly by definition the nonmetricity allows for changes of geodesic distance, and hence the mass. A way to completely eliminate this process is to require the addition rule to be commutative and hence the momentum space to be torsion-free. If, by  similar arguments as in this work, a process could be constructed, in which the momentum change is due to non-associativity, we would then have to exclude curved momentum spaces. But a commutative and associative addition rule is necessarily the special relativistic one with a Minkowski momentum space.

Finally, following on the last point, this could be a genuine physical prediction of the theory - at higher energies one might expect that in some processes a particle might change its total momentum by a loop process. Momentum then would be conserved locally, but not on a global scale. At lower energies, this would be a small effect, but perhaps measurable in some high precision interferometry experiments. Usually in particle physics experiments such a process would be assumed to have been an emission of a particle that was not detected, for example a neutrino. If this were a physical process, we would have to be able to find the difference in signature of this process as opposed to emission of an unobserved particle.

\noindent
{ \bf Acknowledgements:}
 I would like to thank  Giovanni Amelino-Camelia, Lin-Qing Chen, Laurent Freidel, Jerzy Kowalski-Glikman and Lee Smolin  for useful discussions and comments. Research at Perimeter Institute is supported by the Government of Canada through Industry
Canada and by the Province of Ontario through the Ministry of Economic Development \& Innovation.


\end{document}